\documentclass[journal,onecolumn]{IEEEtran}
\ifCLASSINFOpdf
\else
\fi
\usepackage[utf8]{inputenc}
\usepackage[T1]{fontenc}
\usepackage[swedish,english]{babel}
\usepackage{cite}
\usepackage{amsfonts}
\usepackage{amsmath}
\usepackage{amssymb}
\usepackage{mathtools}
\usepackage{mathrsfs}
\usepackage[symbol]{footmisc}

\usepackage{graphicx, float, subfigure, blindtext}
\usepackage{color}

\newcommand{\bs}[1]{\boldsymbol{#1}}

\begin{document}
%
\title{Technical Note:\\ 
PDE-constrained Optimization Formulation for Tumor Growth Model Calibration
}
%
%
%
\author{\IEEEauthorblockN{ %
Baoshan Liang\IEEEauthorrefmark{1},
Luke Lozenski\IEEEauthorrefmark{2},
Umberto Villa\IEEEauthorrefmark{3}, 
Danial Faghihi\IEEEauthorrefmark{1}
}\\
\IEEEauthorblockA{
\IEEEauthorrefmark{1}Mechanical and Aerospace Engineering Department, University at Buffalo\\
\IEEEauthorrefmark{2}Electrical and Systems Engineering Department, Washington University in St. Louis\\
\IEEEauthorrefmark{3}Oden Institute for Computational Engineering and
Sciences, The University of Texas at Austin
}
}
\markboth{PDE-constrained optimization for tumor model calibration}%
{Shell \MakeLowercase{\textit{et al.}}: Bare Demo of IEEEtran.cls for IEEE Journals}
%



\maketitle



%
\IEEEpeerreviewmaketitle
\section{introduction}
\IEEEPARstart{W}{E} discuss solution algorithms for calibrating a tumor growth model using imaging data posed as a deterministic inverse problem. The forward model consists of a nonlinear and time-dependent reaction-diffusion partial differential equation (PDE) with unknown parameters (diffusivity and proliferation rate) being spatial fields. We use a dimension-independent globalized, inexact Newton Conjugate Gradient algorithm to solve the PDE-constrained optimization. The required gradient and Hessian actions are also presented using the adjoint method and Lagrangian formalism.

\section{Tumor Growth Model}\label{section:fwd_model}
\noindent The tumor model employed in this work is the 
standard single-species reaction-diffusion partial differential equation (PDE) \cite{murray2001, harpold2007}, capturing tumor proliferation and infiltration,
\begin{eqnarray}\label{eq:forward}
\frac{\partial u}{\partial t} = 
\nabla \cdot \left( D \nabla {u} \right) + 
G \left( 1 - {u} \right) u
& \text{in} & \Omega \times (0, T], \nonumber \\
D \nabla {u} \cdot \bs{n} = 0
& \text{on} & \Gamma \times (0, T], \nonumber \\
u (\bs{x}, 0) = u_0
& \text{in} & \Omega .
\end{eqnarray}
%
Here, $\Gamma$ is the domain boundary with outward unit normal $\bs{n}$ corresponds to the inner surface of the brain and is not infiltrated by the tumor cells.
The bounded domain $\Omega$ is the brain of a particular rat, segmented from a 2D slice of $T_2$-weighted MRI of each rat,  and the state variable, 
$u(\bs{x},t) \in \mathcal{V}=L^2(0,T; H^1_0(\Omega))\bigcap L^\infty(0,T; H^1_0(\Omega))$ 
is the model estimated tumor volume fraction at each spatial point $\bs{x}$ in the brain domain $\Omega$ and time domain $t\in(0,T]$ with values ranging from 0 to 1, where $T$ is the final time in days.
Parameter ${D} (\bs{x}) \in \tilde{\mathcal{M}}$ accounts for the tumor spread in the host tissue due to invasion and cell migration, and
$G(\bs{x}) \in \tilde{\mathcal{M}}$ denotes the proliferation rate of the tumor, the rate at which tumor cells grow and divide, where $\tilde{\mathcal{M}}\subseteq L^2(\Omega)$. Consequently, the unknown parameters vector is defined as $\boldsymbol \theta (x)=  \begin{pmatrix} D(x) \\ G(x) \end{pmatrix} \in \tilde{\mathcal{M}} \times \tilde{\mathcal{M}} = \mathcal{M}$.
%
%
The unknown model parameters are subject-specific, and spatially varying to account for the brain tissue and tumor heterogeneities. 
Here, the model parameters represent the logarithm of the tumor diffusion and proliferation coefficients, as this ensures the positivity of these estimated quantities in the subsequent Bayesian inversion. 

\section{Deterministic inverse problem as PDE-constrained optimization problem}\label{section:inverse}
\noindent Measurements $d$ and parameters ${\bs \theta} = (D,G)$ are associated via the relationship

\begin{equation}\label{eq:setup1}
    d = \mathcal{F}(\bs \theta) + \eta, 
\end{equation}
where $\mathcal{F}: \mathcal{M} \mapsto \mathcal{V}$ is the \textit{parameter-to-observable map}, which maps the parameters $D$ and $G$ to tumor spatial-temporal distribution $u$, and $\eta$ represent measurement noise. Here, we assume that the data $d$, obtained from MRI,  can also be represented as a function in $\mathcal{V}$. 
Specifically, the parameter to observable map in equation \ref{eq:setup1} can be written as:
\begin{equation}\label{eqn:newsetup}
    \mathcal{F}({\bs \theta}) = \mathcal{B}u({\bs \theta}), \quad \text{ where } u({\bs \theta}) \text{ solves }r(u, {\bs \theta}) = 0.
\end{equation}
Above, the observation operator $\mathcal{B}: \mathcal{V} \mapsto \mathcal{V}$ is a sum of spatiotemporal Dirac deltas corresponding to the observation points and times, and the forward model $r(u, {\bs \theta}) = 0$ is given by \eqref{eq:forward}.
%
%
%
The goal of an inverse problem is then to estimate the unknown parameters $D,G$ given measurements $d$ and the parameter-to-observable map $\mathcal{F}$.
Inverse problems of this form are ill-posed and difficult to solve because small changes measurements can result in large changes in reconstructed parameters \cite{tikhonov1977}. To address this issue, we introduce Tikhonov regularization on the deterministic inverse problem, which penalizes the oscillatory components of the parameters, promotes smooth variation in the solution \cite{Engl1996,Vogel2002}. The deterministic inverse problem can then be formulated as a  regularized least squares optimization problem:
 \begin{equation}\label{eq:objective}
     \underset{D, G\in \mathcal{M}}{\text{argmin}}\mathcal{J}(D, G)
     =
     \underset{D, G \in \mathcal{M}}{\text{argmin}}\left\{\frac{1}{2 \sigma_{noise}^2} \int_0^T \int_\Omega \left[\mathcal{B}(u - d)\right]^2 \;d\bs{x}dt + \mathcal{R}(D, G)\right\}
 \end{equation}
\noindent where $\mathcal{J}(D, G)$ is the cost functional consisting of two terms: the first term is the \textit{data misfit term} weighted by the inverse of the noise covariance $\sigma^{2}_{noise}$, and $\mathcal{R}(D, G)$ is the \textit{regularization term} imposed on the parameter $D, G$ to address the instability issue in the solution of an inverse problem. We defined the regularization term as the following: 
\begin{equation}
    \mathcal{R}(D, G) = \frac{1}{2}\int_\Omega\left[\mathcal{A}_{\rm D}(D - \bar{D})\right]^2\;d\bs{x} + \frac{1}{2}\int_\Omega\left[\mathcal{A}_{\rm G}(G - \bar{G})\right]^2\;d\bs{x}
\end{equation}
\noindent where $\bar{D}$ and $\bar{G}$ are the mean of the prior distributions, and $\mathcal{A}_{\rm D}$, $\mathcal{A}_{\rm G}$ are self-adjoint differential operator stemming from the inverse square root of the parameter covariance priors $\mathcal{C}_{\rm D}$, $\mathcal{C}_{\rm G}$ \cite{bui2013}. 
In particular, for $\gamma_D, \gamma_G > 0$ and $\delta_D, \delta_G \geq 0$, we define 

\begin{equation}
\mathcal{A}_{\rm D} D = \left\{ \begin{array}{ll} 
- \nabla \cdot \left( \gamma_D \nabla D\right) + \delta_D D & \text{ in } \Omega\\
\gamma_D \nabla D \cdot \boldsymbol{n} & \text{ on } \Gamma
\end{array}\right., \text{ and }
\mathcal{A}_{\rm G} G = \left\{ \begin{array}{ll} 
- \nabla \cdot \left( \gamma_G \nabla G\right) + \delta_G G & \text{ in } \Omega\\
\gamma_G \nabla G \cdot \boldsymbol{n} & \text{ on } \Gamma
\end{array}\right.
\end{equation}
%
%

\section{Gradient computation}
\subsection{Lagrangian associated with the PDE-constrained optimization problem}
\noindent The method of Lagrangian is a strategic approach to solving the PDE-constrained optimization problem without explicit parameterization of the constraints, yet the derivatives of the objective function can still be applied. It converts the constrained optimization problem defined in section \ref{section:inverse} into a Lagrangian form collecting the objective function, and the constraints multiplied by \textit{adjoint variables}. Hence, the minimization of the objective functions with the PDE constraints is equivalent to the minimization of the Lagrangian. 
%
%
To this aim, we write the Lagrangian functional of the PDE-constrained optimization problem associated with \ref{eq:forward} and \ref{eq:objective} as:
\begin{equation}\label{eq:lagrangian1}
    \mathscr{L}(u, \{D,G\}, \{p, p_0\}) := \mathcal{J}(D, G) + (p, r(u, D, G)) + (p_0, r_0(u, D, G))
\end{equation}
\noindent where $p\in\mathcal{V}_0$ is the adjoint variable and $p_0\in H_0^1(\Omega)$ is the adjoint variable for the initial condition. The first term is the objective function. The second and the third term are the L2-inner products of $p$ and the residual $r$, and of $p_0$ and the residual at the initial time $r_0$. $(\cdot, \cdot)$ denotes the $L^2$-inner product, i.e. $(u(x), v(x))$ is defined as:
\begin{equation}\label{eq:L2inner}
    (u(x), v(x)) = \int_\Omega u(x)v(x)dx
\end{equation}
To complete the Lagrangian form of the adjoint equation, we need to first derive the weak form of the residual by multiplying with the Lagrange multiplier (or the adjoint variable) and integrating over the space-time domain:

\begin{equation}\label{eq:weakform_residual1}
    \int_0^T \int_\Omega p \left[ 
\frac{\partial u}{\partial t}  
+ \nabla \cdot \left( D \nabla {u} \right) 
- G \left( 1 -  {u} \right) u
\right] \;d\bs{x} dt = 0,
\end{equation}
Further, reduce the equation by applying integration by parts. 
\begin{equation}\label{eq:weakform_residual2}
\int_0^T \int_\Omega \left[ 
p \frac{\partial u}{\partial t}  
+ D \nabla u \cdot  \nabla p  
- p G ( u -  {u}^2 )
\right] \; d\bs{x} dt 
 + \int_0^T \int_\Gamma p D \nabla u \cdot \mathbf{n} \; ds dt 
= 0.
\end{equation}
\noindent As mentioned previously, the adjoint variable $p$ is zero over boundary $\Gamma$, the second integral becomes zero.
Now, we can formulate the full Lagrangian functional as:
\begin{equation}\label{eq:adjLagrangian}\begin{split}
\mathscr{L}(u, \{D,G\}, \{p, p_0\}) :=
     \frac{1}{2 \sigma_{noise}^2} \int_0^T \int_\Omega \left[\mathcal{B}(u - d)\right]^2 \;d\bs{x}dt
+
\frac{1}{2}\int_\Omega\left[\mathcal{A}_{\rm D}(D - \bar{D})\right]^2\;d\bs{x} + \frac{1}{2}\int_\Omega\left[\mathcal{A}_{\rm G}(G - \bar{G})\right]^2\;d\bs{x}\\
+ \int_0^T \int_\Omega \left[ 
p \frac{\partial u}{\partial t}  
+ D \nabla u \cdot  \nabla p  
- p G ( u -  {u}^2 )
\right] \; d\bs{x} dt  
+ \int_\Omega p_0 (u(\bs{x},0) - u_0) \; d\bs{x}
\end{split}\end{equation}
\noindent The variations of the Lagrangian functional $\mathscr{L}(u, \{D,G\}, \{p, p_0\})$ with respect to all variables must vanish and result in 0 to solve the optimization problem.

\subsection{Forward problem: variation of Lagrangian with respect to adjoint variables}\label{section:forward_problem}
\noindent To obtain the adjoint weak form, we need to take the variation of the Lagrangian,
$\mathscr{L}$ with respect to $p$ in an arbitrary direction $\Tilde{p}$, and with respect to $p_0$ in an arbitrary direction $\Tilde{p}_0$, which we denote as $(\delta_{p}\mathscr{L}, \Tilde{p})$, and $(\delta_{p_0}\mathscr{L}, \Tilde{p}_0)$ respectively. 
\begin{equation}\label{eq:forward_problem_p}
(\delta_{p}\mathscr{L}, \Tilde{p}) =
 \int_0^T \int_\Omega \left[ 
\Tilde{p} \frac{\partial u}{\partial t}  
+ D \nabla u \cdot  \nabla \Tilde{p}  
- \Tilde{p} G ( u -  {u}^2 )
\right] \; d\bs{x} dt = 0 \quad\forall \Tilde{p}(\bs{x},t)\in\mathcal{V}_0
\end{equation}
\begin{equation}\label{eq:forward_problem_q}
(\delta_{p_0}\mathscr{L}, \Tilde{p}_0) =
\int_\Omega \Tilde{p}_0 (u(\bs{x},0) - u_0) \; d\bs{x} = 0 \quad\forall \Tilde{p}_0(\bs{x}) = \Tilde{p}(\bs{x},0)\in\mathcal{V}_0
\end{equation}
Then, we can see that the variation of the Lagrangian with respect to $p$ and $q$ gives back the forward problem as its strong form:
\begin{eqnarray}\label{eq:fwd_strong}
    \frac{\partial u}{\partial t} = 
\nabla \cdot \left( D \nabla {u} \right) + 
G \left( 1 - {u} \right) u
& \text{in} & \Omega \times (0,T], \nonumber \\
D \nabla {u} \cdot \mathbf{n} = 0
& \text{on} & \Gamma \times (0,T], \nonumber \\
u (\bs{x}, 0) = u_0
& \text{in} & \Omega .
\end{eqnarray}

\subsection{Adjoint Problem: variation of Lagrangian with respect to state variable}\label{section:adjoint_problem}
\noindent To obtain the adjoint weak form, we need to take the variation of the Lagrangian, $\mathscr{L}$ with respect to $u$. 
\begin{equation}\label{eq:weakform_adjoint}
\begin{split}
     (\delta_{u}\mathscr{L}, \Tilde{u}) = \int_0^T \int_\Omega \frac{1}{\sigma_{noise}^2} \mathcal{B}^*\mathcal{B}(u - d) \Tilde{u} \; d\bs{x}dt
+ \int_0^T \int_\Omega
\left[ 
p \frac{\partial \Tilde{u}}{\partial t}  
+ D \nabla \Tilde{u} \cdot  \nabla {p}  
- p G ( \Tilde{u} -  2u\Tilde{u} )\right] \; d\bs{x} dt \\
+ \int_\Omega p_0 \Tilde{u}(\bs{x},0)  \; d\bs{x} = 0   \quad\forall\Tilde{u}(\bs{x}, t)\in\mathcal{V}_0
\end{split}
\end{equation}
Perform integration by parts, we get the weak form of the adjoint problem, $\delta_u\mathscr{L}$:
\begin{equation}\begin{split}
    (\delta_{u}\mathscr{L}, \Tilde{u}) = \int_0^T \int_\Omega
\Tilde{u}\left[ 
\frac{1}{\sigma_{noise}^2}\mathcal{B}^* \mathcal{B}(u - d)    
- \frac{\partial p}{\partial t}  
- \nabla \cdot (D \nabla {p} )
- p G ( 1 -  2u )\right] \; d\bs{x}dt\\
+ \int_\Omega \Tilde{u}(\bs{x},T) p(\bs{x},T) \; d\bs{x}
+ \int_\Omega \Tilde{u}(\bs{x},0) [p_0 - p(\bs{x},0) ]  \; d\bs{x} = 0 \quad\forall\Tilde{u}(\bs{x}, t)\in\mathcal{V}_0
\end{split}
\end{equation}
Then, the strong form of the adjoint is the following, 
\begin{eqnarray}
- \frac{\partial p}{\partial t}  \nonumber
- \nabla \cdot \left( D \nabla p \right) 
- p G \left( 1 -  2u \right)
&=& - \frac{1}{\sigma_{noise}^2} \mathcal{B}^*\mathcal{B}(u - d) \quad \text{in} \quad  \Omega \times (0,T]  \\\nonumber
D \nabla {p} \cdot \mathbf{n} &=& 0
\quad  \text{on} \quad   \Gamma \times (0,T]   \\\nonumber
p (\bs{x}, T) &=& 0
\quad  \text{in} \quad  \Omega \\
p_0(\bs{x}) &=& p(\bs{x}, 0) \quad  \text{in} \quad  \Omega \\
\vspace{-0.1in}
\end{eqnarray}
\noindent In the adjoint equation above, we denoted the adjoint of $\mathcal{B}$ as $\mathcal{B}^*$.
While the space-time operator $\mathcal{B}$ operates on the state space, extracting value of the state $u$ at the observation times and locations and zeroing out the state at other locations, $\mathcal{B}^*$ does the opposite, going from data space to state space. In general, $\mathcal{B}$ is NOT a self-adjoint operator.
However, in the tumor problem that the data space is defined to be the same as the state space, then $\mathcal{B}^*$ is the same as $\mathcal{B}$: It takes a data function that has non-zero values at observation locations and zeroes otherwise.

\subsection{Expression of Gradient with respect to unknown parameters}
\noindent We take the derivative of $\mathscr{L}$ with respect to $D$ and $G$ to derive the two components of the gradient in the arbitrary directions $\Tilde{D}$, and $\Tilde{G}$. Each gradient evaluation involves solving a pair of forward (equation \ref{eq:forward_problem_p},\ref{eq:forward_problem_q}) and adjoint problems (equation \ref{eq:weakform_adjoint}). In other words, $u$, $p$ solves the forward and adjoint problems, accordingly.
\begin{equation}\label{eq: gradient_D}
(\mathcal{G}_D(D), \Tilde{D}) 
\coloneqq (\delta_{D}\mathscr{L}, \Tilde{D}) = 
\int_0^T \int_\Omega
\Tilde{D} \nabla u \cdot \nabla p \; d\bs{x} dt
+ 
\int_\Omega\mathcal{A}^2_{\rm D} (D - \bar{D}) \,\Tilde{D}\; d\bs{x} \quad\forall \Tilde{D}(\bs{x}) \in\Tilde{\mathcal{M}}
\end{equation}
\begin{equation}\label{eq: gradient_G}
(\mathcal{G}_{G}(G), \Tilde{G}) 
\coloneqq
(\delta_{G}\mathscr{L}, \Tilde{G}) = 
\int_0^T \int_\Omega
-p (u - u^2) \;\Tilde{G} d\bs{x} dt
+ 
\int_\Omega \mathcal{A}^2_{\rm G}(G - \bar{G}) \, \Tilde{G}  \; d\bs{x} \quad\forall \Tilde{G}(\bs{x})\in\Tilde{\mathcal{M}}
\end{equation}
\noindent The strong form of the gradient will give the following. $\mathcal{G}_D$, and $\mathcal{G}_G$ are the gradient of the Lagrangian with respect to $D$ and $G$ correspondingly. 
\begin{equation}\label{eq:strong_gradient_D}
\mathcal{G}_D(D) =
    \begin{cases}
    \mathcal{A}^2_{\rm D}(D - \bar{D}) + \nabla u \cdot \nabla p&\text{in}\quad\Omega\times(0,T]\\
     \gamma_D\nabla \left( \mathcal{A}_D D \right)\cdot\bs{n}&\text{on}\quad\Gamma\times(0,T]
    \end{cases}
\end{equation}
\begin{equation}\label{eq:strong_gradient_G}
\mathcal{G}_G(G) =
    \begin{cases}
    \mathcal{A}^2_{\rm G}(G - \bar{G}) - p (u - u^2) &\text{in}\quad\Omega\times(0,T]\\
     \gamma_G\nabla \left( \mathcal{A}_G G\right)\cdot\bs{n}&\text{on}\quad\Gamma\times(0,T]
    \end{cases}
\end{equation}


\section{Evaluation of Hessian action}
\subsection{Meta-Lagrangian for the evaluation of Hessian action}\label{section: Hessian}
\noindent The main difficulty of this evaluation is that we need to construct Hessian, to be more specific, it is the construction of the misfit Hessian. The number of solutions for the adjoint/forward pair problems is highly dimensional-dependant so we can not construct the Hessian explicitly, and it is also impossible for computers to store and factor such a large matrix. Instead, we can compute the action of Hessian on $\hat{D}$ and $\hat{G}$, i.e. $\mathcal{H}_D\hat{D}$ and $\mathcal{H}_G\hat{G}$. Then, the adjoint-based approach can also be applied to computing the Hessian action. Here, we formulate the computation of Hessian action. We use the a meta-Lagrangian function for the Hessian: 
\begin{equation}\label{eq:Hessian}
\begin{split}
        \mathscr{L}^\mathcal{H}(u,\{D,G\},\{p,p_0\};\hat{u},\{\hat{D},\hat{G}\},\{\hat{p},\hat{p}_0\})\coloneqq(\hat D, \mathcal{G}_D) + (\hat{G}, \mathcal{G}_G) + (\delta_{p}\mathscr{L}, \hat{p}) + (\delta_{p_0}\mathscr{L}, \hat{p}_0) +(\delta_{u}\mathscr{L}, \hat{u})
\end{split}
\end{equation}
\noindent The first two term is the gradients with respect to $D$ and $G$, defined in equation \ref{eq: gradient_D} and \ref{eq: gradient_G}. The third and forth terms are from the forward problem defined in equation \ref{eq:forward_problem_p}, and \ref{eq:forward_problem_q}, and the last term is the adjoint problem defined in equation \ref{eq:weakform_adjoint}. 

\subsection{Incremental forward problem}
\noindent The variation of the Lagrangian w.r.t $p$ and $p_0$ with arbitrary directions $\Tilde{p_0}$ and $\Tilde{p}_0$, let it be called $(\delta_{p}\mathscr{L}^\mathcal{H}, \Tilde{p})$, and $(\delta_{q}\mathscr{L}^\mathcal{H}, \Tilde{q})$.
\begin{equation}\label{eq:incre_fwd1}
\begin{split}
    (\delta_{p}\mathscr{L}^\mathcal{H}, \Tilde{p}) =
    \int_0^T\int_\Omega\left[ \Tilde{p}\frac{\partial\hat{u}}{\partial t} + D\nabla\hat{u}\cdot\nabla\Tilde{p} - \Tilde{p}G\hat{u}(1 - 2u) -\Tilde{p}\hat{G}(u - u^2)\right]\;d\bs{x}dt  = 0 \quad \forall \Tilde{p}(\bs{x}, t)\in\mathcal{V}_0
\end{split}
\end{equation}

\begin{equation}\label{eq:incre_fwd2}
    (\delta_{p_0}\mathscr{L}^\mathcal{H}, \Tilde{p}_0) \coloneqq \int_\Omega\hat{u}(\bs{x},0)\Tilde{p}_0\;d\bs{x} = 0\quad\forall\Tilde{p}_0(\bs{x}) = \Tilde{p}(\bs{x}, 0)\in\mathcal{V}_0
\end{equation}
Then, bringing the weak form defined in equation \ref{eq:incre_fwd1} and \ref{eq:incre_fwd2} to strong form, we get the incremental forward problem as below:
\begin{eqnarray}\label{eq:incre_fwd_strong}
    \frac{\partial \hat{u}}{\partial t} = 
\nabla \cdot \left( D \nabla \hat{u} \right) + \nabla \cdot \left( \hat{D} \nabla {u} \right) \nonumber\\
+ \hat{G} \left( 1 - {u} \right) u + G \left( 1 - {u} \right) \hat{u} \nonumber
&\text{in}& \Omega \times (0,T]  \\
D \nabla \hat{u} \cdot \mathbf{n} = 0
&\text{on}& \Gamma \times (0,T]\\
\hat{u}(\bs{x}, 0) = 0&\text{in}&\Omega.\nonumber
\end{eqnarray}

\subsection{Incremental adjoin problem}
\noindent The variation of the Lagrangian w.r.t the state variable, $u$ in an arbitrary direction of $\Tilde{u}$. Let it be $(\delta_{u}\mathscr{L}^\mathcal{H}, \Tilde{u})$.

\begin{equation}\label{eq;incre_adj}
    \begin{split}
       (\delta_{u}\mathscr{L}^\mathcal{H}, \Tilde{u}) = -\int_0^T\int_\Omega p(\Tilde{u}-2u\Tilde{u})\hat{G}\;d\bs{x}dt + \int_0^T\int_\Omega \hat{D}\nabla\Tilde{u}\cdot\nabla p\;d\bs{x}dt + \int_0^T\int_\Omega\Tilde{u}\left[ \frac{1}{\sigma_{noise}^2} \mathcal{B}^*\mathcal{B}\hat{u} + 2pG\hat{u} \right]\;d\bs{x}dt\\
       + \int_0^T\int_\Omega\left[ \hat{p}\frac{\partial\Tilde{u}}{\partial t} - D\nabla\Tilde{u}\cdot\nabla\hat{p} -\hat{p}G(\Tilde{u} - 2u\Tilde{u}) \right]\;d\bs{x}dt + \int_\Omega\hat{p}_0\Tilde{u}(\bs{x}, 0)\;d\bs{x} = 0\quad\forall\Tilde{u}\in\mathcal{V}_0
    \end{split}
\end{equation}
\noindent We derive the strong form from the weak form defined in equation \ref{eq;incre_adj} as the incremental adjoint problem:

\begin{eqnarray}\label{eq:incre_adj_strong}
\frac{\partial \hat{p}}{\partial t} + \left[ \nabla \cdot \left( \hat{D} \nabla p \right) + \nabla \cdot \left( D \nabla \hat{p} \right) \right] &-& \left[\hat{p} G \left( 1 -  2u \right) + p \hat{G} \left( 1 -  2u \right) +- 2pG\hat{u} \right]\nonumber\\
= - \frac{1}{\sigma_{noise}^2} \mathcal{B}^*\mathcal{B}(\hat{u}) &\text{in}&  \Omega \times (0,T]  \\\nonumber
\hat{D} \nabla {p} \cdot \mathbf{n} + D \nabla \hat{p} \cdot \mathbf{n} = 0
&\text{on}&   \Gamma \times (0,T]   \\\nonumber
p_0 (\bs{x}) = 0
&\text{in}&  \Omega 
\end{eqnarray}

\subsection{Expression of Hessian action}
\noindent To obtain the Hessian action, we take the derivative of the Lagrangian, $\mathscr{L}^H$ defined in equation \ref{eq:Hessian} with respect to the two unknown parameters $D, G$ in the directions of $\Tilde{D}$, $\Tilde{G}$ respectively.
\begin{equation}
\begin{split}
    (\Tilde{D}, \mathcal{H}_{DD}\hat{D}+\mathcal{H}_{GD}\hat{G}) \coloneqq (\delta_{D} \mathscr{L}^\mathcal{H}, \Tilde{D}) = \int_\Omega\Tilde{D}\mathcal{A}^2_{\rm D}\hat{D}\;d\bs{x} + \int_0^T\int_\Omega\Tilde{D}\nabla u\cdot\nabla\hat{p}\;d\bs{x}dt 
    \\
    + \int_0^T\int_\Omega\hat{u}\nabla\cdot(\Tilde{D}\nabla p)\;d\bs{x}dt = 0 \quad\forall\Tilde{D}(\bs{x})\in\Tilde{\mathcal{M}}
\end{split}
\end{equation}

\noindent $(\Tilde{D}, \mathcal{H}_{DD}\hat{D}+\mathcal{H}_{GD}\hat{G})$ denotes the variation of the Hessian action in a direction of $\hat{D}$. $u$ is the solution to the forward problem defined in equations \ref{eq:forward_problem_p} and \ref{eq:forward_problem_q}. $p$ and $q$ are the solutions to the adjoint problem in equation \ref{eq:weakform_adjoint}. $\hat{u}$ is the incremental state, the solution to the \textit{incremental forward problem}. $\hat{p}$ and $\hat{q}$ is the incremental adjoint, the solution to the \textit{incremental adjoint problems}. The incremental forward and adjoint problems are obtained by setting to zero variations of $\mathscr{L}^H$ with respect to $p$, $q$, and $u$, respectively, defined in \ref{eq:incre_fwd1}, \ref{eq:incre_fwd2}, and \ref{eq;incre_adj}. Similarly, we can derive the $G$ component of the Hessian action as:
\begin{equation}
    (\Tilde{G}, \mathcal{H}_{GG}\hat{G}+\mathcal{H}_{DG}\hat{D}) \coloneqq (\delta_{G} \mathscr{L}^\mathcal{H}, \Tilde{G}) = \int_\Omega\Tilde{G}\mathcal{A}^2_{\rm G}\hat{G}\;d\bs{x} - \int_0^T\int_\Omega\hat{p}\Tilde{G}(u - u^2)\;d\bs{x}dt - \int_0^T\int_\Omega\hat{u}p\Tilde{G}(1 - 2u)\;d\bs{x}dt = 0 \quad\forall\Tilde{G}(\bs{x})\in\Tilde{\mathcal{M}}.
\end{equation}
\noindent Then, we can obtain the strong form of the Hessian action: 
\begin{equation}\label{eq:strong_Hessianaction_D}
\mathcal{H}_{DD}\hat{D}+\mathcal{H}_{GD}\hat{G} = 
    \begin{cases}
        \mathcal{A}^2_{\rm D}\hat{D} + \left( \nabla \hat{u} \cdot \nabla p + \nabla u \cdot \nabla \hat{p}\right)&\text{in}\quad \Omega \times (0,T]  \\
        \gamma_D\nabla\left( \mathcal{A}_D \hat{D} \right)\cdot\bs{n}&\text{on}\quad  \Gamma \times (0,T]
    \end{cases}
\end{equation}
\begin{equation}\label{eq:strong_Hessianaction_G}
\mathcal{H}_{GG}\hat{G}+\mathcal{H}_{DG}\hat{D} =
    \begin{cases}
       \mathcal{A}^2_{\rm G}\hat{G} -  \left[(]\hat{p} (u - u^2) + p\hat{u}(1 - 2u)\right]&\text{in}\quad \Omega \times (0,T]  \\
        \gamma_G\nabla\left( \mathcal{A}_G \hat{G} \right)\cdot\bs{n}&\text{on}\quad  \Gamma \times (0,T]
    \end{cases}
\end{equation}
\section{Model Verification}
\begin{figure}[h!]
    \centering
    \includegraphics[width=0.5\linewidth]{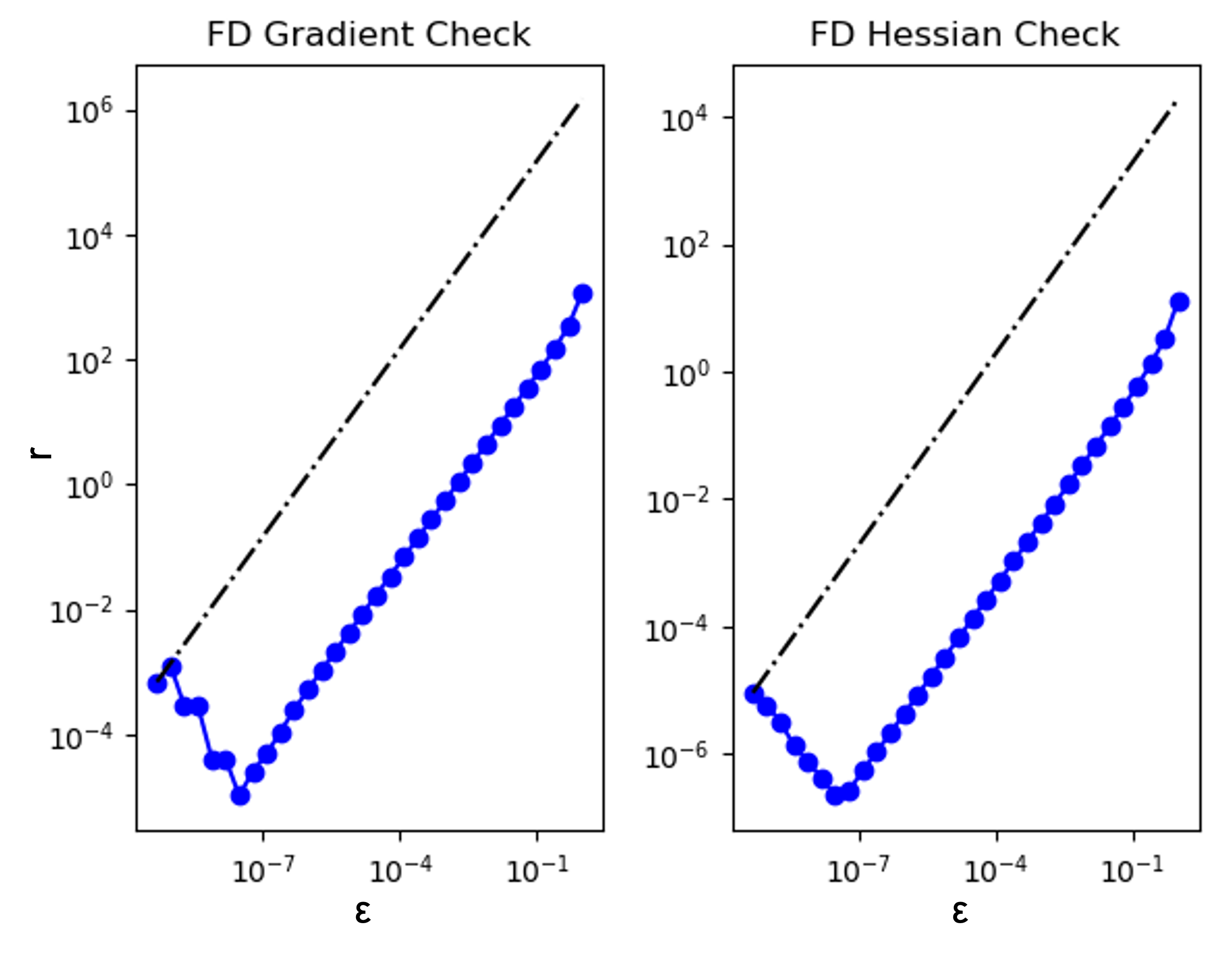}
    \vspace{-10pt}    
    \caption{The black dotted lines are the theoretically largest r values for gradient (left-hand side panel) and for Hessian (right-hand side panel) for each step size. The blue dotted lines are the actual r for gradient (left-hand side panel) and in Hessian (right-hand side panel) given different $\varepsilon$.}
    \label{fig:modelVerify}
\end{figure}

\noindent When implementing the above formulation, we perform finite element discretization \textit{in space} with piecewise quadratic globally continuous Galerkin elements, and finite difference discretization \textit{in time} with implicit Euler to the weak form of the forward, adjoint, and incremental problems. The parameter $\theta = (D,G)$ is discretized using piecewise linear globally continuous Lagrangian finite elements. The expression of gradients and Hessian actions are implemented in \texttt{FEniCS}\cite{fenics} and \texttt{hIPPYlib}\cite{hippylib}. 
A finite difference check is performed to verify the derivation and implementation of the expressions above. 
Specifically, given a point $\bs{\theta}_0 \in \mathcal{M}$ in parameter space and an arbitrary (randomly sampled) direction $\Tilde{\bs{\theta}}\in \mathcal{M}$, the finite difference check for the gradient reads
\begin{equation}
    r := \left| \frac{ \mathcal{J}(\bs{\theta}_0 + \varepsilon \tilde{\bs{\theta}}) - \mathcal{J}(\bs{\theta}_0)}{\varepsilon} -  \left(\mathcal{G}(\bs{\theta}_0), \tilde{\bs{\theta}}\right)\right| = \mathcal{O}(\varepsilon)\,\end{equation}

\noindent where $\mathcal{G}(\bs{\theta}_0)$ denotes the gradient with respect to $\bs{\theta}$, evaluated at $\bs{\theta}_0$. Similarly, we can check the Hessian with the following:
\begin{equation}
   r := \left\| \frac{ \mathcal{G}(\bs{\theta}_0 + \varepsilon \hat{\bs{\theta}}) - \mathcal{G}(\bs{\theta}_0)}{\varepsilon} -  \mathcal{H}_0 \hat{\bs{\theta}}\right\| = \mathcal{O}(\varepsilon),
\end{equation}
\noindent where $\mathcal{H}_0$ is the Hessian action evaluated at $\bs{\theta}_0$. In Figure. \ref{fig:modelVerify}, we show the value of $r$ as a function of $\varepsilon$ in a log-log scale. We observe that $r$ decays linearly for a wide range of values of $\varepsilon$. We note the increases in the errors for extremely small values of $\varepsilon$ due to numerical stability and finite precision arithmetic.







\vspace{-10pt}

\ifCLASSOPTIONcaptionsoff
  \newpage
\fi

\newpage
\bibliographystyle{IEEEtran}
\bibliography{bibtex/bib/reference}

\end{document}